%% file: paper-01-v2-arxiv.tex
\documentclass[runningheads]{llncs}

\usepackage[T1]{fontenc}
\usepackage{booktabs}
\usepackage{amsmath}
\usepackage{array}
\usepackage{graphicx}

\begin{document}

\title{Population-Calibrated Graph Screening at 835-Million-Address Scale, with Label-Free Transfer to New Chains}
\titlerunning{Population-Calibrated Graph Screening at 835M-Address Scale}

\author{Yury Korolev}
\authorrunning{Y. Korolev}
\institute{AIDECISIONS, London}

\maketitle

\begin{abstract}
Compliance screening of blockchain addresses is, in practice, a lookup against sanctions registries plus clustering heuristics; it fails on unlabelled addresses and on chains with no label coverage at all. We describe a deployed system that scores an address by its position in a multi-chain transaction graph rather than by its presence in a list. The substrate is a single graph of \textbf{835,330,427} addresses and \textbf{15,826,261,934} edges across five EVM chains; a shared inductive encoder with per-chain normalisation feeds two scoring heads. Decision thresholds are \textbf{exact quantiles of the score distribution over the full population}, scanned per chain segment, so the alert volume is known in advance. We report: \textbf{label-free transfer} --- heads trained on two chains recall \textbf{0.8598 / 0.8182 / 0.9967} of held-out positives on Base, Arbitrum and Gnosis at a $10^{-3}$ population alert rate, with no target-chain labels in head training; a \textbf{static lead-time replay} over 68 external registry events --- \textbf{40 of 68 (58.8\,\%)} flagged at the 0.1\,\% budget, $\times$152 over an event-level random-flagging baseline, with first on-chain appearance a median of \textbf{528.8 days} (Ethereum) / 647.8 days (Tron) before public designation; a \textbf{serving path} whose score is bit-identical to the offline artefact at end-to-end p50 \textbf{151\,ms}, gated by a 2,882-address drift panel; and an \textbf{adversarial harness} of eight recurrent reinforcement-learned archetypes that passes an 8-criterion degeneracy audit and, on a detector-independent snapshot, exposes a measured blind spot of the deployed heads against synthesised behaviour.

\end{abstract}

\section{Introduction}

An address arrives at a compliance desk. The question is not ``is this address on a list'' --- that is a solved database join --- but whether the address \emph{behaves} like the addresses on the list. Production screening stacks answer it with registry lookups, co-spend clustering, and vendor attribution of uninspectable provenance. Three failure modes are structural rather than incidental.

\textbf{Label sparsity is the binding constraint, not model capacity.} Public designation lists are small, lag the activity they describe by months to years, and are distributed unevenly across chains. A model that can only speak about labelled addresses inherits that distribution. We quantify the marginal value of labels directly (Sect.~\ref{sec:leakage}): doubling the labelled pool buys +0.018 recall on Ethereum and +0.014 on Tron --- a real but modest return, and the empirical argument for putting the signal in the graph rather than in the list.

\textbf{Coverage is per-chain, but crime is not.} Funds cross bridges in minutes; a stack with a model for Ethereum and nothing for the destination chain has no answer when it matters, and the standard remedy --- collect labels for the new chain, then train --- reintroduces the first constraint. Sect.~\ref{sec:transfer} measures the alternative: apply existing heads to a chain that contributed no labels to training.

\textbf{An alert budget you cannot state in advance is not deployable.} A threshold chosen to optimise a validation metric tells an operator nothing about how many alerts arrive on Monday. Over 323 million Ethereum addresses, the difference between a threshold estimated from a negative sample and one measured over the population is roughly one histogram bin --- thousands of alerts at a $10^{-3}$ budget. Sect.~\ref{sec:calibration} replaces estimation with an exact population scan.

\textbf{Scope.} This paper is about the five-chain EVM model: substrate, calibration, transfer, evaluation, serving, and the adversarial harness we probe it with. The deployed product also screens Bitcoin through a separate label-based path; it is out of scope and no claim here covers it.

\textbf{What we open and what we keep.} This paper is published alongside an open-source release of the \emph{label-production tooling} it depends on: the address-label schema and parsers, registry scrapers for nine jurisdictions, the VASP, mixer and bridge registry builders, and the tag-pack generator, the adversarial simulator of Sect.~\ref{sec:harness}, and the detector's training code --- architecture, warm-start recipe, head training and the threshold-calibration mechanism, and the evaluation harness of the lead-time replay in Sect.~\ref{sec:leadtime} --- replay assembly, the event-level null and lift, the publication-date audit and the audit-table emitter (\texttt{github.com/ai-decisions/openlabels}, \texttt{github.com/ai-decisions/openabm}, \texttt{github.com/ai-decisions/openheads}, \texttt{github.com/ai-decisions/openeval}). We do \textbf{not} release the trained weights, the substrate, the emitted embeddings, the address and provenance indices, or the serving stack. The line is deliberate: open is the \textbf{instrument}, closed is the \textbf{result of applying it to our own data}. Sect.~\ref{sec:leakage} quantifies the closed side --- doubling the labelled pool moves recall by +0.018, so the tooling alone does not reconstitute the system. Label claims are third-party verifiable by running the released tooling on the same public inputs (sha256 of the emitted pack); the model's figures are auditable rather than reproducible (Appendix~\ref{app:repro}).

\textbf{Contributions.}

\begin{enumerate}
\item A single multi-chain substrate that places the screening signal in an inductive graph representation rather than in target-chain labels: 835,330,427 addresses $\times$ 25 features, 15,826,261,934 edges, five EVM chains in one node space with contiguous per-chain index segments (Sect.~\ref{sec:substrate}).
\item Calibration by exact full-population scan, with the complete threshold grid, realised flagged counts and two measured pathologies that make nominal alert-rate points non-comparable across chains (Sect.~\ref{sec:calibration}).
\item A label-free transfer result under an explicit protocol that states, per component, what did and did not use target-chain data (Sect.~\ref{sec:transfer}).
\item A lead-time replay against 68 external registry events with a stated static-replay convention and an event-level random baseline, reporting detection rate and lead jointly (Sect.~\ref{sec:leadtime}).
\item A production path with bit-exactness as an acceptance criterion, honest degradation for unserved chains, and a drift panel that gates model swaps (Sect.~\ref{sec:serving}).
\item A reinforcement-learning adversarial harness for this class of detector, with an ablation that contradicts our own pre-specified expectation about why it works (Sect.~\ref{sec:harness}).
\end{enumerate}

We deliberately do not claim: cross-chain entity resolution (L2/L3); completeness of the autonomous-agent class beyond one chain (L5); or any improvement of the detector obtained by adversarial co-training (out of the claim set, Sect.~\ref{sec:harness}).

\section{Related work}

\textbf{Supervised detection on public blockchain data.} Graph learning for anti-money laundering was framed as a scalability problem before it was a benchmark one~\cite{weber2018scalable}; the Elliptic Bitcoin dataset~\cite{weber2019} is a widely used benchmark for illicit-transaction classification --- its original evaluation uses a temporal split --- with convolutional~\cite{alarab2020} and dynamic~\cite{pareja2020evolvegcn} architectures evaluated on it. The line has since moved to account-based chains and specific crime types --- graph embeddings~\cite{wu2022phishers} and edge-temporal aggregation~\cite{li2022ttagn} for Ethereum phishing --- to label scarcity via self-supervised embeddings~\cite{lo2023inspection}, to larger ground truth: Bitcoin forensics~\cite{elmougy2023forensics} and the subgraph-level Elliptic2~\cite{bellei2024elliptic2}, and, into 2025--2026, to graph-neural AML classification~\cite{ferretti2025aml}, dynamic and heterogeneous-graph Ethereum-phishing architectures~\cite{li2026dynamicphish,wang2026pdhg} and hierarchical graph transformers for fraud detection~\cite{gao2026ahgt}. Expressivity results for directed multigraphs~\cite{egressy2024ppgn} treat the motifs laundering produces; synthetic generators~\cite{altman2023aml} supply labelled behaviour where real labels are absent.

Our work sits alongside that line rather than inside it, and the difference is the unit of evaluation. Those results --- the 2025--2026 architectures included --- are reported on one chain, against curated labelled datasets, at operating points chosen on the evaluation sample. We report on five chains in one node space, at operating points measured over the full deployed population, against ground truth external to the label pipeline. We do \emph{not} report a result on the standard public benchmark: our own recorded Elliptic run does not satisfy the protocol discipline we require of the numbers in this paper --- no validation split, and an F1 computed at $k = n_{\mathrm{positives}}$ rather than at a threshold, hence not comparable to the published threshold-F1 baselines --- and we state its absence rather than ship a number that would not survive a reviewer's rerun (Sect.~\ref{sec:limitations}, L12).

\textbf{Inductive graph representation learning at serving scale.} Our encoder follows the neighbour-sampling inductive family~\cite{hamilton2017}, with attention~\cite{velickovic2018gat} and heterogeneous~\cite{hu2020hgt} variants as the adjacent design space; that choice is what makes serving possible: embeddings for 835M nodes are materialised once and a new address is scored by lookup, not by retraining. Web-scale deployments~\cite{ying2018pinsage} and partition- and subgraph-sampling training regimes~\cite{chiang2019clustergcn,zeng2020graphsaint} establish that graphs of this size are trainable; the engineering consequences of \emph{serving} one are under-reported --- Sect.~\ref{sec:model} and Sect.~\ref{sec:serving} describe them.

\textbf{Blockchain forensics and measurement.} Full-graph measurement~\cite{ron2013quantitative} and address-clustering heuristics~\cite{androulaki2013privacy,meiklejohn2013} are the attribution primitives of this field; end-to-end tracing of a criminal revenue pipeline~\cite{huang2018ransomware} is the precedent for evaluating against real-world events rather than held-out labels --- the design of Sect.~\ref{sec:leadtime}. Cross-ledger tracing~\cite{yousaf2019tracing} and mixer measurement~\cite{wang2023mixers} address the multi-chain and obfuscation surfaces. We report a \emph{failed} attempt at cross-chain entity resolution on our own embeddings (Sect.~\ref{sec:limitations}, L2/L3): a node space that is block-diagonal by construction cannot support cross-chain identity claims regardless of encoder quality --- our transfer result (Sect.~\ref{sec:transfer}) applies a model to a new chain; it does not link identities across chains.

\textbf{Autonomous agents as a screening class.} A substantial share of on-chain activity is generated by programs competing for extractable value~\cite{daian2020flashboys,qin2022bev}; that population is why our second head exists --- an address may be automated without being illicit, and a surface that cannot separate the two produces alerts an operator cannot action; our audit of the class (Sect.~\ref{sec:audit}) is reported against our own prior.

\textbf{Adversarial evaluation and simulation.} Adversary models in security measurement are increasingly economic rather than gradient-based~\cite{apruzzese2023real}, and agent-based simulators are an established way to evaluate controls against generated financial behaviour~\cite{lopezrojas2018paysim,altman2023aml}. Our policies follow the recurrent-policy POMDP line~\cite{hausknecht2015}, trained with PPO~\cite{schulman2017}; the specific construction --- separate actor and critic recurrence, observation augmented with the previous action, entropy regularisation against premature deterministic collapse --- is ours, and the use of historical opponent snapshots echoes league-style multi-agent training~\cite{vinyals2019,berner2019dota}. Sect.~\ref{sec:harness} reports which of those choices mattered, by ablation.

\textbf{Calibration and evaluation discipline.} Reporting recall at a fixed operating point is standard; computing that point over the deployed population is not. Neural classifiers are miscalibrated in ways that make a score threshold a poor proxy for an alert volume~\cite{guo2017calibration}, and the security-measurement literature documents how temporal and sampling bias inflate reported performance~\cite{pendlebury2019tesseract} and catalogues the pitfalls a deployed evaluation must avoid~\cite{arp2022dosdonts}. Those pitfalls are why Sect.~\ref{sec:calibration} replaces an estimated threshold with an exact population scan, why Sect.~\ref{sec:leadtime} states its static-replay convention rather than claiming a backtest, and why Sect.~\ref{sec:harness} treats the evaluation snapshot as part of the threat model.

\section{Substrate}
\label{sec:substrate}

\subsection{What is in the graph}

The substrate is a single node space over five chains, built so that a node identifier is globally unique and chain membership is a contiguous segment of the index. This matters for both normalisation (Sect.~\ref{sec:encoder}) and calibration (Sect.~\ref{sec:calibration}).

\begin{table}[!htbp]
\centering
\caption{Substrate properties. Per-chain segment sizes are taken from the full-population threshold scan and sum exactly to the node count --- a consistency check a reviewer can perform on the numbers as printed.}
\label{tab:substrate}
\begin{tabular}{lr}
\toprule
Addresses (nodes), 5 chains in one node space & \textbf{835,330,427} \\
Node features & \textbf{25} \\
Edges & \textbf{15,826,261,934} \\
\midrule
Ethereum segment & 323,139,851 \\
Tron segment & 331,982,925 \\
Base segment & 131,409,131 \\
Arbitrum segment & 39,134,190 \\
Gnosis segment & 9,664,330 \\
\bottomrule
\end{tabular}
\end{table}

Node features are structural and behavioural aggregates over an address's transaction history: counts, value statistics, temporal spread, counterparty diversity, contract-interaction ratios; Sect.~\ref{sec:limitations} (L4) records the history of a dead-column measurement on an earlier feature epoch and why no effective-dimensionality claim is carried for the current 25-feature line.

\subsection{Labels}
\label{sec:labels}

The training label set (v8) holds \textbf{301,304} rows with provenance recorded for \textbf{100\,\%}; the unified serving label store (v5) holds \textbf{202,369} keys, \textbf{0} outside schema. The class \texttt{ai\_agent} contributes \textbf{89,499} v8 rows (Base 63,372 $\cdot$ Ethereum 16,430 $\cdot$ six other EVM spaces 9,697), with 15,191 of the 16,430 Ethereum rows in-graph. Labels come from public designation lists, public attribution repositories and our own harvesting of public sources; provenance is stored with every row --- what makes the class audit of Sect.~\ref{sec:audit} possible.

\section{Model}
\label{sec:model}

\subsection{Task and metric}

Two heads answer two questions over one embedding: \texttt{fincrime} (does this address behave like designated criminal infrastructure) and \texttt{ai\_agent} (is this address an autonomous economic agent --- a class largely absent from public taxonomies).

The product metric over a union pool is \textbf{either-head recall at a fixed population alert rate}: each positive is scored by \emph{its own} head and the joint alert rate is calibrated empirically. This is not cosmetic: a single head over a union pool understates recall by construction --- it is asked about nodes the other head is trained to reject; an earlier iteration measured \textbf{0.44} single-head against \textbf{0.596} either-head on the same checkpoint, 42\,\% of the union pool being the other head's nodes. All recall figures below are either-head unless a table says \texttt{fincrime} or \texttt{ai\_agent}.

\subsection{Encoder}
\label{sec:encoder}

The trunk is a three-layer GraphSAGE~\cite{hamilton2017} with mean aggregation over the 25 input features --- batch normalisation and ReLU after each convolution, dropout 0.3 after the first two, width 128 throughout, no residual connections. Normalisation is \textbf{per chain}: one learnable LayerNorm per segment on the output embedding, routed by node index against the segment boundaries (one routing subtlety this creates at scale, and its acceptance gate, are described in Appendix~\ref{app:training}). The heads are small MLPs over the shared embedding --- \texttt{ai\_agent} $128{\to}64{\to}1$, \texttt{fincrime} $128{\to}256{\to}128{\to}1$, ReLU and dropout 0.3 between layers --- with a sigmoid at serve time.

\subsection{Training}
\label{sec:training}

Warm start with a strict full load of the previous backbone (25$\to$25 features), heads freshly initialised at framework defaults; Adam without weight decay, learning rate $5\cdot10^{-4}$ / $10^{-3}$ (backbone / heads); batch 2,048; two-hop neighbour sampling, fanout [25, 10]; 3 epochs; seed 42; splits verbatim from the previous iteration so the comparison in Sect.~\ref{sec:comparison} is like-for-like; loss composition, split protocol and the early-stop rule are stated in Appendix~\ref{app:training}. Validation trajectory across the three epochs: \textbf{0.9088 $\to$ 0.9220 $\to$ 0.9303}, no early stopping.

Two decisions constrain the claims. The 5,222 token-tail \texttt{ai\_agent} labels were \textbf{excluded} from training, because including them would have changed the recipe against which the baselines are defined. And only the warm-start arm was run --- there is no from-scratch arm, so we do not claim warm-starting was necessary, only that it was sufficient.

Training ran in 81.1--82.3 minutes on one 8$\times$A100 node. Full-graph embedding emission ran in 122.4--124.7 minutes on a second identical node, producing 1,671 chunks of 500K rows each (fp16, 128 dims): \textbf{213,848,356,544} bytes, which equals $835{,}330{,}427 \times 128 \times 2$ plus chunk headers, uploaded in 11 minutes. A serving claim over 835M nodes is only credible if re-emission is routine rather than heroic.

\subsection{Calibration by exact population scan}
\label{sec:calibration}

Thresholds are \textbf{population quantiles measured by scanning every score in the graph}, per chain segment. The scan is exact, not sampled; it closes a gap in the previous iteration, where $\tau$ was estimated from a negative sample --- the two differ by about one bin ($\approx$1.35\,\% of the count) --- and both are now recorded side by side whenever a threshold moves between score carriers.

A note on terminology, because it is load-bearing: the quantile controls the \textbf{population alert rate} (a screening budget), \emph{not} a false-positive rate --- the population contains an unknown number of true positives, so the false share of the flagged 0.1\,\% is not identified. We write ``recall at population alert rate $10^{-3}$'' throughout and reserve ``FPR'' for pools with known negatives; alert volume is what an operator plans against, and it is exactly what the quantile fixes.

The complete grid, with the realised flagged counts:

\begin{table}[!htbp]
\centering
\caption{Complete \texttt{fincrime}-score threshold grid with realised flagged counts.}
\label{tab:taugrid}
\setlength{\tabcolsep}{2pt}
\scriptsize
\resizebox{\textwidth}{!}{%
\begin{tabular}{lrccc}
\toprule
Chain & n in segment & $\tau$ @$10^{-2}$ (flagged) & $\tau$ @$10^{-3}$ (flagged) & $\tau$ @$10^{-4}$ (flagged) \\
\midrule
Ethereum & 323,139,851 & 0.6144 (3,231,393) & \textbf{0.7968} (323,140) & 0.9286 (32,313) \\
Tron & 331,982,925 & 0.5946 (3,319,828) & \textbf{0.7464} (331,982) & 0.8477 (33,199) \\
Base & 131,409,131 & 0.5754 (1,314,090) & \textbf{0.8690} (131,410) & 0.9778 (281) \\
Arbitrum & 39,134,190 & 0.5630 (391,342) & \textbf{0.8315} (39,135) & 0.9359 (3,914) \\
Gnosis & 9,664,330 & 0.0486 (96,644) & \textbf{0.5431} (9,665) & 0.9690 (967) \\
\bottomrule
\end{tabular}%
}
\end{table}

Two measured properties follow from publishing the counts; both are findings, not defects of the scan.

The Gnosis threshold at $10^{-2}$ is \textbf{0.0486}, far below every other chain: the score distribution on that chain is compressed toward zero, so a 1\,\% population budget reaches deep into the bulk. And the Base flagged count at $10^{-4}$ is \textbf{281} where a strict $10^{-4}$ of 131M would be $\approx$13,141: the quantile lands on a plateau of tied scores, so the \emph{realised} alert rate at that point is far below the nominal one.

The consequence generalises beyond our system: \textbf{``recall at alert rate $10^{-4}$'' is not comparable across chains unless the flagged counts are printed next to it.} Any cross-chain table of recall at a nominal rate --- ours included --- must be read with the realised counts in hand; every recall table in this paper therefore carries its threshold and pool sizes.

\section{Label-free transfer to new chains}
\label{sec:transfer}

Base, Arbitrum and Gnosis contributed no labels to head training. ``Label-free'' is a per-component statement, and we make it per component rather than calling the chains ``unseen'': the \textbf{heads} saw no positives or negatives from these chains; the shared \textbf{encoder} was trained on the merged graph and therefore did see the target chains' \emph{structure}, unsupervised; the per-chain \textbf{normalisation} uses each target chain's population statistics; and the decision \textbf{threshold} is the target chain's own population quantile (Sect.~\ref{sec:calibration}). What transfers with no target-chain input is the supervised component; everything the protocol consumes from the target chain is unlabelled. The evaluation holds out 20\,\% of each chain's labelled positives, samples 500,000 negatives per chain, and applies the population $\tau$ --- the same thresholds a production user is served at.

\begin{table}[!htbp]
\centering
\caption{Label-free transfer at the served thresholds.}
\label{tab:transfer}
\begin{tabular}{lcc}
\toprule
Chain & Recall @ population alert rate $10^{-3}$ & $\tau$ @$10^{-3}$ \\
\midrule
Base & \textbf{0.8598} & 0.8690 \\
Arbitrum & \textbf{0.8182} & 0.8315 \\
Gnosis & \textbf{0.9967} & 0.5431 \\
\bottomrule
\end{tabular}
\end{table}

The claim we make is that the representation transfers: heads trained with no labels from these chains recall 82--100\,\% of held-out positives there at a 0.1\,\% population alert budget.

The claim we do \textbf{not} make concerns Gnosis specifically. Its 0.9967 must be read together with its threshold (0.5431, the lowest in the grid) and its compressed score distribution (Sect.~\ref{sec:calibration}): a compressed distribution can produce very high recall at a threshold that is low in absolute terms. We report it as measured and flag it as the least informative of the three.

The operational reading: a chain can be brought onto the screening surface without first acquiring labels for it.

\section{Evaluation}
\label{sec:evaluation}

\subsection{Per-chain recall grid at population thresholds}

Evaluation pools are per-chain samples of the graph with all known positives retained. Thresholds are the population $\tau$ of Sect.~\ref{sec:calibration}, so a recall figure here is directly the recall a production user gets at that population alert budget. These all-positive pools differ from the 20\,\%-held-out protocol of Sect.~\ref{sec:transfer} --- hence Base 0.8494 here against 0.8598 there: same thresholds, different pools.

\begin{table}[!htbp]
\centering
\caption{Per-chain recall at population thresholds (recall at population alert rate $10^{-2}$ / $10^{-3}$ / $10^{-4}$).}
\label{tab:recallgrid}
\setlength{\tabcolsep}{2pt}
\scriptsize
\resizebox{\textwidth}{!}{%
\begin{tabular}{lrrrcc}
\toprule
Chain & Pool nodes & fincrime pos. & ai\_agent pos. & fincrime & either-head \\
\midrule
Ethereum & 2,522,621 & 12,656 & 10,007 & 0.9421 / \textbf{0.8537} / 0.6688 & 0.9610 / \textbf{0.8427} / 0.5035 \\
Tron & 2,501,482 & 1,482 & 0 & 0.7659 / \textbf{0.5931} / 0.5027 & --- (single head) \\
Base & 563,432 & 63,432 & 0 & 0.9843 / \textbf{0.8494} / 0.3697 & --- \\
Arbitrum & 500,829 & 829 & 0 & 0.9831 / \textbf{0.8721} / 0.7117 & --- \\
Gnosis & 501,525 & 1,525 & 0 & 1.0000 / \textbf{0.9980} / 0.3541 & --- \\
\bottomrule
\end{tabular}%
}
\end{table}

The Ethereum \texttt{ai\_agent} head alone: 0.9772 / 0.7853 / 0.2881 at $\tau_{\mathrm{ai}}$ 0.6503 / 0.8837 / 0.9636.

Note that either-head recall on Ethereum at $10^{-3}$ (0.8427) is \emph{below} the fincrime head alone on the fincrime subset (0.8537): the union pool adds 10,007 agent-class positives whose head is weaker at that threshold --- the honest reading of a two-head system, and why we publish both columns.

\subsection{Headline pooled figures}
\label{sec:pooled}

On the standing evaluation pool, 5-fold pooled, either-head recall at population alert rate $10^{-3}$: Ethereum \textbf{0.8314} ($\sigma$ 0.008 across folds, $n = 22{,}621$); Tron \textbf{0.5783} ($\sigma$ 0.048, $n = 1{,}482$).

\subsection{Comparison against the previous production line}
\label{sec:comparison}

Identical protocol, splits, seed and folds, previous $\to$ current: Ethereum 0.7607 $\to$ \textbf{0.8314} (+0.0707); Tron 0.5655 $\to$ \textbf{0.5783} (+0.0128); Base 0.8518 $\to$ \textbf{0.8598} (+0.0080); Arbitrum 0.7818 $\to$ \textbf{0.8182} (+0.0364); Gnosis 0.9934 $\to$ \textbf{0.9967} (+0.0033). Five of five chains improved; none regressed.

A controlled feature-only baseline (same features, labels and split; own population quantiles) reaches 0.837 leak-free test recall at $10^{-3}$ against the graph's 0.9289, and falls 0.2--0.3 behind on Base and Arbitrum (Appendix~\ref{app:ablation}).

\subsection{Leakage control and label scaling}
\label{sec:leakage}

A multi-label node can be held out for one head while remaining a training positive for the other. That is a real leakage channel and a na\"ive split does not close it: we check the held-out set against \textbf{every} training pool that contributes positives, not only the primary one. Under that control, test-split recall is \textbf{0.9289}, against the validation trajectory of Sect.~\ref{sec:training}.

Doubling the labelled pool is worth \textbf{+0.018} recall on Ethereum and \textbf{+0.014} on Tron, measured across 5 folds in the earlier embedding space. Two consequences: label acquisition has a modest, measurable marginal return at our pool size --- the empirical form of Sect.~1's argument for structural signal over label accumulation; and the Tron gap in Sect.~\ref{sec:pooled} is label coverage, not architecture (supported independently at L1).

\subsection{Lead-time replay against external registry events}
\label{sec:leadtime}

We take registry events (designations, hacks, exchange collapses) with addresses on Ethereum and Tron, score every address, flag when the \texttt{fincrime} score meets its chain's population $\tau$ at $10^{-3}$, and compare each address's \textbf{first on-chain appearance} with the event's public publication date. First-seen block times come from the substrate itself (one scan per chain), not a third-party API, so the replay input is reproducible from the model's own data. An event counts as detected when at least one of its in-graph addresses is flagged. Because an event carries 1--9 in-graph addresses (median 4), the correct random baseline is \textbf{event-level}: flagging 0.1\,\% of addresses uniformly at random detects an event with $m$ scoreable addresses with probability $1-(1-10^{-3})^{m}$. The null-rate column is the mean of that quantity over the group's events, and the lift is the detection rate against it; under this null the expected number of detected events in the EVM total is \textbf{0.26 of 68}, against 40 observed.

\begin{table}[!htbp]
\centering
\caption{Lead-time replay at the \texttt{fincrime}-head $10^{-3}$ population threshold, with the event-level null.}
\label{tab:leadtime}
\setlength{\tabcolsep}{2pt}
\scriptsize
\resizebox{\textwidth}{!}{%
\begin{tabular}{lrrrrrrr}
\toprule
Group & Events & Detected & Rate & Null rate & Lift & Median lead (d) & Pos.\ lead \\
\midrule
Ethereum & 42 & 23 & 54.8\,\% & 0.394\,\% & $\times$139 & \textbf{528.8} & 20 \\
Tron & 26 & 17 & 65.4\,\% & 0.376\,\% & $\times$174 & \textbf{647.8} & 17 \\
\textbf{EVM total} & \textbf{68} & \textbf{40} & \textbf{58.8\,\%} & \textbf{0.387\,\%} & \textbf{$\times$152} & \textbf{578.4} & 37 \\
Scraped publication dates only & 58 & 33 & 56.9\,\% & 0.389\,\% & $\times$146 & \textbf{647.8} & 31 \\
\bottomrule
\end{tabular}%
}
\end{table}

\textbf{The convention, stated so it cannot be misread.} This is a \emph{static replay}, and it establishes a retrospective association, not an early warning. The features the heads score are aggregates over the address's \textbf{complete} history, including activity after its first appearance, so the replay does \textbf{not} show that the signal was extractable at first appearance, and it is not a backtest of a deployed system. The claim it supports is narrower: addresses that today's model flags at the 0.1\,\% budget first appeared on chain a median of 528.8 days (Ethereum) / 647.8 days (Tron) before the event's public designation. The measurement it would take to say more --- history truncated at time $T$, features and embedding rebuilt at $T$, scored at $T$ --- is a time-censored replay we have not run (Sect.~\ref{sec:limitations}, L6).

Two properties give the replay its evidential weight. The ground truth is \textbf{external} to our label and evaluation pipeline --- registry events, not held-out labels --- and the threshold is the population quantile, so the comparison is against a stated alert budget rather than a balanced pool. In operational terms: flagging 1 address in 1,000 recovers more than half of the registry events on those two chains, where random flagging at the same budget would be expected to recover 0.26 of the 68.

Detection rate matters as much as lead, and we report them jointly rather than quoting the lead alone: 37 of the 40 detections precede publication, and the audited-date subset --- 58 of 68 events after the registry date audit --- detects 33 at a median lead of 647.8 days. Appendix~\ref{app:events} lists all 68 events with dates, provenance (including the audit that replaced 28 registry default dates with the publishing OFAC action), per-event counts and the inclusion criteria.

\subsection{Audit of the autonomous-agent class}
\label{sec:audit}

The \texttt{ai\_agent} class is our own construction, so it requires an audit a reviewer can check. Of 10,007 agent-class positives in the Ethereum pool, \textbf{all 10,007 are explained by a label row} --- behavioural-only, unexplained positives: \textbf{0}. The curated source line that originally motivated the class (``yield contracts are economic agents'') contributed \textbf{0} rows to the training label set; only 8 specific addresses from that hypothesis reached training, through an older store epoch, and all 8 sit in the pool. Their share of the pool's agent positives is \textbf{0.0008} (0.08\,\%).

Verdict: the class is clean to within 8 addresses. Its largest actual contributors are protocol-level tags --- DEX routers, MEV bots, AVS operators, vault strategies --- consistent with the class definition fixed in advance (a label seed plus a behavioural filter) and \emph{not} with the yield-contract hypothesis that motivated it. Reporting the disagreement with our own prior is what makes the audit meaningful.

\section{Serving}
\label{sec:serving}

\subsection{The path an address takes}

A screening call resolves the address to a node index through an 835,330,427-key index, reads its 128-dimensional embedding from the materialised store, evaluates both heads, and compares against the per-chain $\tau$. Nothing is recomputed at request time --- the graph is untouched, there is no neighbour sampling --- which is what makes both the latency and the bit-exactness gate below achievable. Measured latency, two vantages: end-to-end authenticated HTTPS (network and API tier included) p50 \textbf{151\,ms} / p95 256\,ms ($n=15$, one warm address); server-side on the scoring tier over $n=5{,}000$ calls cycling 144 registry addresses, p50 \textbf{175.9} / p95 188.1 / p99 922.5\,ms --- first touch of an address p50 636.7\,ms, minimum observed 0.9\,ms: the spread is the embedding store's chunk cache, so latency depends on address locality (L8).

\subsection{Bit-exactness and honest degradation as acceptance gates}
\label{sec:bitexact}

A production score is accepted only if it equals the offline artefact exactly. For a well-known mixer contract the deployed path returns \textbf{0.8933565020561218}, bit-identical to the offline computation, under model version \texttt{63.v75-5chain-\allowbreak d465968d17fde8ce}; the model version is returned in every response.

Any model swap is additionally gated by a \textbf{2,882-address drift panel} across the three transfer chains: every address must reproduce its recorded score with $\Delta = 0$ against the canonical stored artefact. During the adversarial campaign of Sect.~\ref{sec:harness} the panel returned exact zeros on all three chains --- how we know that work did not perturb the deployed line.

Degradation is explicit rather than silent. For a chain outside the served line the API returns a null score with an explicit \texttt{detector\_unserved\_chain} flag rather than a default value; if the scoring tier is unreachable, the response carries a degradation flag rather than a fabricated score. ``Never return a number you cannot reproduce'' is the design rule; both flags exist so it survives partial failure.

\section{An adversarial harness for this class of detector}
\label{sec:harness}

\textbf{Why and setup.} A detector's evaluation pool is drawn from the same distribution as its labels, which says nothing about behaviour a determined adversary would produce; we therefore train adversaries, let them act, and report what the harness establishes. Eight adversary archetypes --- validator compromiser, infrastructure compromiser, entity-lineage evader, regulatory arbitrageur, mixer launderer, CEX-offramp seeker, evolving mixer, agent coordinator --- are trained with PPO~\cite{schulman2017} in a simulator whose world state is built from real chain snapshots. Policies are recurrent (separate actor and critic recurrence; previous-action observation; entropy regularisation; Sect.~2). The simulator is released as open source: 26 modules, 13,064 lines of package and script code, 430 tests.

\textbf{Result 1 --- the policies are non-degenerate.} An 8-criterion reward-hacking and degeneracy audit passes \textbf{8/8} on two independent runs and 7/8 on a third, where the only failing criterion is single-rejection-code concentration (0.72 against a 0.70 threshold). The action entropy of the three archetypes that previously collapsed to a single action is $> 0$ \textbf{on all 13 checkpoints}. A non-degenerate adversary is the precondition for every other statement in this section.

\textbf{Result 2 --- what fixed the collapse, by ablation, against our own expectation.} Our hypothesis was that collapse came from an uninformative observation (a constant neighbour embedding) and that real structural embeddings plus recurrence would fix it. An ablation with the embeddings \textbf{zeroed} still passes 7/8: the fix is the \textbf{recurrence}, not the structural embeddings --- reported because it contradicts our pre-specified expectation, and the cheaper ingredient is the one that matters.

\textbf{Result 3 --- long-horizon stability.} At a rollout horizon four times longer than the audit configuration (604--621 steps per archetype) there is no collapse: action entropy 1.270--1.889, dominant-action share $\le 0.65$, amount mean 0.31--0.55 with $\sigma$ 0.21--0.24. The one criterion that concentrates at long horizon is a single \emph{environment} rejection code (bridge liquidity exceeded, 0.77--0.81, on 4 of 4 runs for the regulatory arbitrageur; slippage, 0.70--0.72, on 2 of 4 for the agent coordinator). That is a property of the simulated environment's constraint surface, not of the policy --- the entropy figures above come from the same runs.

\textbf{Result 4 --- evaluation snapshots are part of the threat model.} Scoring ``the addresses our adversary touched'' against a baseline is invalid unless the snapshot is drawn \textbf{independently of the model under test}. We ran the same evaluation twice: once seeded from a detector-\textbf{stratified} snapshot whose starting addresses were the top-100 highest-scoring addresses in the graph (baseline mean 0.2729, p95 0.8635), once from a \textbf{uniformly random} snapshot ($n = 2{,}000$, in-graph coverage 0.8894; baseline mean 0.1092, p95 0.4257).

The two runs return \textbf{opposite verdicts for all eight archetypes}. On the stratified snapshot every archetype reads as detected; on the clean random snapshot every archetype reads as a blind spot --- 8/8, with the largest elevation over background being $\Delta = +0.0366$ against a baseline mean of 0.1092, below the +0.05 threshold fixed in advance. We state the corrected verdict rather than only the fact of the inversion: on a snapshot drawn independently of the detector, \textbf{the deployed heads do not separate the generated activity from background.}

The mechanism is the confound, not the adversary's sophistication: when agents begin inside a detector-hot subgraph and route value to recipients drawn from the same pool, elevated scores measure \emph{where the simulation was confined}, not \emph{what it generated}. The confound was recorded as a caveat before the rerun.

What this licenses: a screening surface calibrated on registry-derived labels has structural blind spots against generated behaviour, and the harness locates them. What it does \textbf{not} license: any weakening of Sect.~\ref{sec:leadtime}, which is measured against 68 \emph{real} registry events on a disjoint support. A model can be effective against the crime that actually occurs and blind to behaviour synthesised outside its training distribution; conflating the two is the error the two snapshots make visible.

\textbf{Fingerprint-level reading.} Alongside score-level verdicts we match a 78-dimensional structural fingerprint per generated bundle by cosine against reference bundles (best-match p50 0.9486--0.9779). The reference set is the simulator's own later generations, not the detector's training subgraph (bundle-level fingerprints of the address-level training pools do not exist), so the similarity is not evidence about the detector's training set; we record the protocol adaptation as such.

\textbf{What this section does not claim.} We do not claim that co-evolving the detector against the simulator improves the detector (a measurement we do not report here; no claim depends on it), and we do not claim the simulated economy models any real market --- its role is to generate structurally plausible adversary behaviour, not to forecast it.

\section{Limitations}
\label{sec:limitations}

Each item is a measurement, not a hedge.

\textbf{L1 --- Tron is our weakest chain and the ceiling is public labels.} Either-head recall 0.5783 at population alert rate $10^{-3}$ against 0.8314 on Ethereum, on a pool of only 1,482 positives ($\sigma$ 0.048 across folds, itself a small-sample artefact). We attribute the gap to label coverage on the evidence of Sect.~\ref{sec:leakage}, not to the backbone; the feature-only baseline reads the same way --- gradient boosting matches the graph on Tron (Appendix~\ref{app:ablation}).

\textbf{L2 --- Cross-chain entity resolution does not work on these embeddings.} Cluster purity 0.410 on Ethereum and 0.333 on Tron against a 0.90 acceptance floor; the cross-chain share of resolved pairs is \textbf{0.000} (0 of 168 candidate pairs). The cause is structural --- see L3.

\textbf{L3 --- The multi-chain graph is block-diagonal by construction.} The earlier two-chain merge contained \textbf{726} cross-chain edges out of 10,728,037,368; chain centroids sit at cosine 0.24; cross-chain retrieval reaches 7.2\,\% at rank 100. The current five-chain merge contains 953 cross-chain links and nothing has been trained on it. Stated as a boundary: we have a per-chain model that transfers (Sect.~\ref{sec:transfer}); we do \textbf{not} have a cross-chain identity capability.

\textbf{L4 --- The dead-column census predates the current feature line, so no effective-dimensionality claim is carried.} On an \emph{earlier} 20-feature input epoch, 7 features were measured identically zero on Base, Arbitrum and Gnosis. The current line uses 25 raw features, and after that census a set of aggregate features was backfilled on exactly those chains; the census has \textbf{not} been repeated on the 25-feature matrix the reported numbers are computed from. We therefore state the transfer results of Sect.~\ref{sec:transfer} on the 25-feature line as built and make no claim, in either direction, about dead columns on the target chains; repeating the census on the current matrix is an open measurement.

\textbf{L5 --- The autonomous-agent class has an evaluation pool on one chain only.} All 10,007 agent-class pool positives are Ethereum. The class has label rows on other chains (Sect.~\ref{sec:labels}) but no evaluation pool there, so every \texttt{ai\_agent} figure in this paper is an Ethereum figure.

\textbf{L6 --- The lead-time replay is static} (Sect.~\ref{sec:leadtime}). Restated because it is the figure most likely to be over-read: 528.8 days median lead on Ethereum is not ``our system warned 529 days early''. It is ``addresses the current model flags --- on features aggregated over their complete history --- first appeared on chain 529 days before public designation, in the 23 of 42 Ethereum events where the model flags an address at all''. Whether the signal was extractable at first appearance is not established; that requires a time-censored replay (features truncated at $T$), which we have not run.

\textbf{L7 --- Threshold comparability across chains.} Nominal alert-rate points do not realise equally across chains because of score ties and distribution compression (Sect.~\ref{sec:calibration}); cross-chain recall at a nominal rate must be read with the flagged counts.

\textbf{L8 --- Latency vantages.} The end-to-end sample is small ($n=15$, one address); the $n=5{,}000$ server-side distribution excludes network and the API tier and is chunk-cache-bound, not compute-bound; read together.

\textbf{L9 --- Full-graph clustering has never been run.} The 506M-node clustering pass required for graph-wide community structure has not completed: the index build exceeded single-device memory (a 259\,GB payload against 80\,GB), and multi-device sharding is not implemented. Community-level statements are therefore local, not global.

\textbf{L10 --- Retrieval is measured but not served.} A quantised approximate-nearest-neighbour index (IVF65536(HNSW32) + SQfp16) over $686{,}363{,}853 \times 128$ vectors supports ``addresses that look like this one'': canary recall@15 \textbf{0.9748}, labelled-retrieval quality \textbf{0.7394} against a 0.5884 baseline. Two qualifications disqualify a capability claim: the index lags the substrate by one epoch (686M vectors against 835M nodes), and the retrieval endpoint is \textbf{not} in the serving path.

\textbf{L11 --- Scope excludes the Bitcoin path.} The deployed product also screens Bitcoin through a separate label-based path with different semantics. It is out of scope: no figure or claim here covers it, and the five-chain results do not extend to it.

\textbf{L12 --- No public benchmark result is reported} (Sect.~2). Our recorded Elliptic run does not meet this paper's protocol standard; we state the absence rather than publish a figure whose statistic differs from the baselines it would be compared against.

\textbf{L13 --- Bounds on Sect.~\ref{sec:harness}.} The harness results are about the simulator and about evaluation methodology. Long-horizon rejection-code concentration is a property of the simulated environment's constraints, not of the policy; the fingerprint reference set is the simulator's own generations rather than the detector's training subgraph.

\textbf{L14 --- The deployed heads have a measured blind spot against synthesised behaviour} (Sect.~\ref{sec:harness}, Result 4): on a snapshot drawn independently of the detector, all eight archetypes score at or below background (8/8, largest $\Delta = +0.0366$ against a baseline mean of 0.1092). Scope: measured on \emph{generated} activity from policies optimised in a simulator, on a support disjoint from the registry events of Sect.~\ref{sec:leadtime}; it does not bear on the 58.8\,\% detection rate there. An operator sizing coverage needs both numbers; the same measurement under a detector-stratified snapshot returns the opposite answer --- why the snapshot protocol is published with the verdict.

\textbf{L15 --- The self-supervised recipe is worse than the supervised one:} 0.559--0.570 against 0.642. We report it because it closes off an obvious ``just pretrain on the unlabelled graph'' reading of our label-sparsity argument. We tried it; at this scale it lost.

\section{Ethics, dual use and data provenance}
\label{sec:ethics}

\textbf{Data.} All chain data is public. Labels derive from public designation lists, public attribution repositories and our own harvesting of public sources; provenance is recorded for 100\,\% of the training label set (Sect.~\ref{sec:labels}). We process on-chain addresses and do not attempt to link them to natural persons or to combine them with off-chain identity data; no such linkage is claimed or released. No human subjects, user studies or personal-data collection were involved; no IRB review was applicable.

\textbf{Dual use.} The adversarial component (Sect.~\ref{sec:harness}) trains policies that generate transaction patterns --- dual-use by nature. We release the \emph{detector-side} and label-side artefacts needed to reproduce our measurements (Appendix~\ref{app:repro}) and do \textbf{not} release adversary policy checkpoints, the simulator's environment configuration, or the generated bundles; Sect.~\ref{sec:harness} is reported at the level of aggregate audit criteria, ablation outcomes and archetype descriptions --- sufficient for the methodological claim, insufficient as an evasion recipe.

\textbf{Responsible disclosure.} The findings of Sect.~\ref{sec:harness} concern our own system and evaluation methodology; no third-party system, protocol or vendor was tested, so there is nothing to disclose to an external party.

\textbf{Deployment harms.} A false positive in this domain can freeze a legitimate user's funds. Two design consequences are load-bearing: thresholds are expressed as population budgets so an operator knows the alert volume in advance (Sect.~\ref{sec:calibration}), and label-based verdicts take precedence over model scores at serve time for registry classes.

\section{Conclusion}

We built the screening surface around the graph rather than around the list. Three findings carry the weight of the paper. First, \textbf{label-free transfer}: heads trained with no labels from Base, Arbitrum or Gnosis recall 0.8598 / 0.8182 / 0.9967 of held-out positives there at a 0.1\,\% population alert budget --- a chain joins the screening surface before any labels exist for it. Second, \textbf{population-calibrated screening}: thresholds are exact quantiles over all 835,330,427 addresses, so the alert volume is stated in advance, and the published grid records where nominal rates do not realise. Third, \textbf{retrospective replay against external ground truth}: 40 of 68 registry events on Ethereum and Tron are flagged at the same budget --- a $\times$152 lift over the event-level random baseline --- with first on-chain appearance a median of 528.8 days before public designation on Ethereum (static-replay convention, Sect.~\ref{sec:leadtime}).

Two further results: the serving path returns a verdict bit-identical to the offline artefact at p50 151\,ms, model version pinned in every response; and the adversarial harness, whose contribution is methodological --- its paired-snapshot experiment shows an adversarial evaluation is interpretable only when the snapshot is drawn independently of the model under test. The system does not provide cross-chain identity (L2, L3), full-graph community structure (L9), or served retrieval (L10).

\appendix

\section{Reproducibility}
\label{app:repro}

\subsection{Label-tooling path (third-party runnable, released)}

The label pipeline emits a standard-format tag pack from public sources: registry scrapers for nine jurisdictions, VASP and mixer/bridge registry builders, and the tag-pack generator. A clean clone run against the same public inputs reproduces the same output, verified by sha256 of the emitted pack.

This tooling is \textbf{released as open source under the Apache License 2.0} at \texttt{github.com/ai-decisions/openlabels}, \texttt{github.com/ai-decisions/openabm}, \texttt{github.com/ai-decisions/openheads} and \texttt{github.com/ai-decisions/openeval}, together with a \texttt{NOTICE} file carrying the source registries' own licence terms and attributions, and a README section stating the open/closed boundary described in Sect.~1. Apache-2.0 rather than a permissive licence without a patent clause: in a segment where the adjacent commercial vendors hold patent portfolios, the express patent grant protects contributors as well as us. Two deliberate exclusions inside the release: regulator register \emph{dumps} are not redistributed where the source's terms do not clearly permit it (only the tooling that fetches them is released), and no data derived from third-party label aggregators is included. Reproducibility for a third party is therefore demonstrated on a public sanctions subset rather than on our full label set.

\subsection{Internal-artefact path (auditable, not publicly runnable)}

Every figure in Sect.~3--8 is produced by one command from the internal repository against artefacts pinned by sha256; the producer script and the sha manifest are committed. This does \textbf{not} make those figures third-party reproducible --- the substrate is hundreds of gigabytes of embeddings over a graph we built --- and we do not claim it does. It makes them \emph{auditable}: a reviewer with access reruns one command and compares byte for byte.

\subsection{What a reviewer can check without our data}

(i) The tag-pack path in A.1, end to end. (ii) Internal consistency of the reported grids: the per-chain segment sizes in Sect.~3.1 sum exactly to 835,330,427; the flagged counts in Sect.~4.4 equal the stated alert rate times the segment size except at the two pathologies flagged explicitly; the embedding store size 213,848,356,544\,B equals $835{,}330{,}427 \times 128 \times 2$ plus chunk headers. (iii) The arithmetic of Sect.~6.5: detection rates are recomputable from the counts as printed, and the lifts from the rates and the printed event-level null rates.

\subsection{What we do not release}

Trained weights, the substrate, embedding emissions, address and provenance indices, the serving stack, and every adversary-side artefact of Sect.~\ref{sec:harness}. Sect.~\ref{sec:ethics} states the reasoning.

\subsection{Training details}
\label{app:training}

The joint loss is $0.3\,\mathrm{BCE}_{\mathrm{ai}} + 0.7\,\mathrm{BCE}_{\mathrm{fincrime}}$, each head's positive weight set to the measured negative-to-positive ratio of its fit split (clamped below at 1); there is no negative mining --- the labelled pool's own negatives carry that role. The split is stratified at seed 42 into three groups (fincrime-positive / agent-only-positive / negative), with 10\,\% frozen test and 10\,\% of the remainder as validation. The early-stop rule --- halt if validation recall at the $10^{-3}$ operating point degrades by more than 0.01 against the previous epoch --- never triggered. The third convolution of the trunk operates within the two-hop sampled subgraph. Every figure here is read from the committed training script, not reconstructed.

One routing detail is invisible until the graph grows: when the Ethereum segment acquired a token-derived tail at indices [686,363,853 \ldots\ 835,330,427), the boundary-bucketing logic routed that tail into the \emph{last} chain's normalisation block. The fix keeps the parameter set byte-identical (the tail boundary is a plain attribute, not a learned tensor), so a strict full-state load from the previous checkpoint remains exact --- itself an acceptance gate, verified on synthetic data before any GPU time was spent.

\section{Number provenance}
\label{app:provenance}

No figure in this paper was produced while writing it. Every figure was frozen before drafting, in a table that binds each number to the artefact that produced it and to that artefact's sha256; the producer script is committed alongside. Derived quantities --- detection rates, lifts --- are recomputed from the frozen counts by script rather than carried by hand, and they are the only figures in the paper that are not read directly off an artefact.

Table~\ref{tab:provenance} states, for each group of figures in the text, \emph{which kind of measurement} produced it. We give the class of artefact rather than our internal identifiers, because an identifier a reader cannot resolve conveys no provenance; what a reader can use is knowing whether a figure came from a full-population scan, a held-out evaluation pool, a live production call, or a simulator audit --- and therefore what it can and cannot support. The internal mapping (frozen-table row identifiers and sha256 manifest) exists for audit under access, as described in Appendix~\ref{app:repro}.

\begin{table}[!htbp]
\centering
\caption{Provenance class of every figure group in the paper.}
\label{tab:provenance}
\scriptsize
\begin{tabular}{p{6.4cm}p{5.0cm}}
\toprule
Figure(s) in text & Produced by \\
\midrule
835,330,427 nodes $\cdot$ 25 features $\cdot$ 15,826,261,934 edges & Substrate build manifest (frozen, sha256) \\
Per-chain segment sizes; complete $\tau$ grid and flagged counts & Exact full-population threshold scan, one pass per chain segment \\
Pooled recall: eth 0.8314 ($\sigma$ 0.008, n=22,621); tron 0.5783 ($\sigma$ 0.048, n=1,482) & 5-fold evaluation on the standing pool \\
Transfer recall: base 0.8598 / arbitrum 0.8182 / gnosis 0.9967 & Held-out transfer protocol (Sect.~\ref{sec:transfer}) at population $\tau$ \\
Previous-line comparison and deltas & Re-run of the previous checkpoint under identical protocol, splits, seed and folds \\
Leak-free test split 0.9289; validation 0.9088 $\to$ 0.9220 $\to$ 0.9303 & Training-run record with the multi-pool leakage control of Sect.~\ref{sec:leakage} \\
Label scaling +0.018 eth / +0.014 tron & Label-ablation sweep, 5 folds \\
Label store sizes: 301,304 rows (provenance 100\,\%); 202,369 serving keys & Label-store audit at build time \\
Retrieval 686,363,853 $\times$ 128; canary recall@15 0.9748; labelled 0.7394 (baseline 0.5884) & Offline index-build and retrieval evaluation (not served) \\
Drift panel 2,882 addresses, $\Delta = 0$ on all chains & Drift-panel comparison against the canonical stored scores \\
Recall grid, all chains and alert-rate points; pool sizes and positive counts & Per-chain evaluation pools at population $\tau$ \\
Agent-class audit: 89,499 / 16,430 / 15,191; curated 0; 8 of 8; share 0.0008 & Provenance join of pool positives against label rows \\
Lead-time per chain and subset: 42/23/528.8/20; 26/17/647.8/17; 58/33/647.8/31 & Registry-event replay against first-seen index and publication dates, after the OFAC date audit of Appendix~\ref{app:events} \\
EVM total 68/40/58.8\,\%; per-group rates, event-level null rates and lifts ($\times$152 EVM; $\times$139 / $\times$174 / $\times$146) & \emph{Derived} from the replay counts and per-event in-graph address counts; recomputed by script \\
Production score 0.8933565020561218 bit-exact; p50 151 / p95 256\,ms (n=15) & Live authenticated calls against the deployed API \\
Server-side latency n=5,000 / 144 addresses (p50 175.9 / p95 188.1 / p99 922.5\,ms; first-touch p50 636.7) & Sequential single-in-flight bench on the scoring tier; producer script and artefact committed \\
Emission 1,671 chunks; 213,848,356,544\,B; train 81.1--82.3 min; emit 122.4--124.7 min & Training and emission run logs \\
Either-head vs.\ single-head 0.44 $\to$ 0.596; 42\,\% of union pool & Metric-comparison record from an earlier checkpoint on the same pool \\
Simulator 26 modules / 13,064 lines / 430 tests; training code 16 modules / 4,671 lines / 116 tests; evaluation harness 9 modules / 1,506 lines / 83 tests & Counted in the published repositories; the test suites run in their CI on Python 3.11 and 3.12 \\
Audit 8/8 twice, 7/8 once (0.72 against a 0.70 threshold); entropy $> 0$ on 13 checkpoints & Degeneracy-audit artefacts, one per checkpoint \\
Zero-embedding ablation 7/8 & Degeneracy audit on the ablated observation \\
Long-horizon 604--621 steps; entropy 1.270--1.889; dominant $\le$ 0.65; amount $\mu$ 0.31--0.55 / $\sigma$ 0.21--0.24; rejection 0.77--0.81 (4/4), 0.70--0.72 (2/4) & Extended-horizon rollout audits \\
Random snapshot n=2,000, coverage 0.8894; baseline mean 0.1092 / p95 0.4257 & Detector-independent snapshot evaluation \\
Stratified snapshot baseline mean 0.2729 / p95 0.8635; opposite verdicts & Detector-stratified snapshot evaluation (the confounded arm) \\
Fingerprint 78-dim, best-match p50 0.9486--0.9779 & Bundle-fingerprint matching artefact \\
Limitations figures (L1--L15) & The measurement that established each limitation, recorded when it was found \\
\bottomrule
\end{tabular}
\end{table}

\textbf{What was withdrawn while preparing this paper, and why.} Two groups of numbers that an earlier draft carried are absent here, and we state the reason rather than letting the absence pass unremarked. A public-benchmark result was withdrawn because its protocol is not defensible: the reported F1 is a top-$k$ statistic at $k = n_{\mathrm{positives}}$ rather than a threshold F1, there is no validation split and the reported AUC is an intermediate-epoch test AUC, and the producer script was not retained (Sect.~\ref{sec:limitations}, L12). A single-chain result outside the five-chain line was withdrawn from the claim set after the served-population measurement showed the capability it asserted does not hold at the operating points we serve; that chain is out of scope here (Sect.~\ref{sec:limitations}, L11). Both measurements are retained internally as product decisions. We report these withdrawals because a paper whose numbers were frozen before drafting should also be explicit about which frozen numbers did not survive review.

\section{The external-event registry: audit of the 68-event replay}
\label{app:events}

Sect.~\ref{sec:leadtime} reports aggregates; this appendix lists every event and states how the set was constructed, so the replay can be audited rather than trusted.

\textbf{Inclusion and exclusion.} The source registry holds 266 public events --- OFAC and other sanctions designations, bridge and DeFi exploits, exchange breaches, key compromises --- each recorded with its public source. An event enters the replay if and only if (i) the registry row carries at least one on-chain address and (ii) at least one of its address forms is on a chain the substrate indexes. 86 events fail (i): the registry records the event but no address list. 2 events fail (ii): their address forms are only on chains outside the substrate. Both criteria are mechanical, fixed before scoring, and independent of model output --- no event was excluded on the basis of what the detector said about it. Of the remaining 178 events, those whose chain is Ethereum (42) or Tron (26) form the 68-event EVM replay; the 110 Bitcoin events are out of scope with the rest of the Bitcoin path (L11). Chain assignment of an address is by its \emph{form}, not the event's chain field: sanctions rows carry multi-chain address lists, and an EVM-hex form is resolved by which segment of the graph holds it.

\textbf{Publication dates and the date audit.} 30 of the 68 events carried a publication date scraped from the public source at replay time (23 Ethereum, 7 Tron); the remaining 38 used a proxy. A date audit of the proxy rows found that 28 of them (11 Ethereum, 17 Tron) sat on the registry's \emph{default} event date of 2020-01-01 --- unreliable by construction. For each of the 28 we resolved the date of the OFAC action that published the entry's digital-currency addresses (designation, or the SDN-entry update that added the addresses --- for entities such as Ansarallah or Cheil Credit Bank the two differ by years), verified against the OFAC recent-actions pages themselves; per-event source URLs ship with the committed date patch. Detection and lift are date-independent and did not move. The lead statistics in Sect.~\ref{sec:leadtime} use the audited dates (Ethereum median 528.8 days, Tron 647.8); under the pre-audit defaults they read 424.5 / 307.8 --- the audit moved the medians \emph{up}. 10 events remain on an event-date proxy; Tables~\ref{tab:events-ethereum} and \ref{tab:events-tron} mark each event's date source.

\textbf{Flagging head.} Flagging in the replay is single-head: an address is flagged when its \texttt{fincrime} score meets its chain's population $\tau$ at $10^{-3}$. The either-head variant (adding the \texttt{ai\_agent} head at its own Ethereum threshold) detects one more event --- 41 of 68 --- with the Ethereum median lead at 269.4 days over 24 detected events; Tron is unchanged. We report the single-head variant throughout because the \texttt{ai\_agent} threshold exists for Ethereum only (Sect.~\ref{sec:limitations}, L5).

\input{event_table_appendix}

\section{Feature-only baselines}
\label{app:ablation}

Whether the graph contributes signal beyond the 25 engineered features is a question a controlled baseline answers. Three non-graph models were trained on the same 25 raw features, the same labelled pool and the same seed-42 split as the production heads (\texttt{fincrime} target): logistic regression on standardised features; gradient boosting (hist, 400 trees, depth 8, positive-class weight the measured 81.5 negative-to-positive ratio of the fit split, early stop on the validation split); and a $25{\to}256{\to}256{\to}1$ MLP mirroring the training discipline of Appendix~\ref{app:training}. Each was then thresholded at population quantiles of \emph{its own} score, computed by the same exact per-chain scan as Sect.~\ref{sec:calibration} --- one full pass over all 835,330,427 addresses per model.

\begin{table}[!htbp]
\centering
\caption{Feature-only baselines against the graph line. ``Test'' = leak-free test-split recall at FPR $10^{-3}$ (ROC convention of Sect.~\ref{sec:leakage}); per-chain columns = \texttt{fincrime} recall on the evaluation pool at each model's own population $\tau$ at alert rate $10^{-3}$ (the grid convention of Table~\ref{tab:recallgrid}; fit rows included for every model equally).}
\label{tab:ablation}
\begin{tabular}{lcccccc}
\toprule
Model & Test & eth & tron & base & arbitrum & gnosis \\
\midrule
Logistic regression & 0.062 & 0.067 & 0.021 & 0.013 & 0.051 & 0.466 \\
Gradient boosting & 0.837 & 0.831 & \textbf{0.594} & 0.544 & 0.657 & 0.947 \\
MLP & 0.605 & 0.418 & 0.070 & 0.418 & 0.493 & 0.722 \\
Graph (production line) & \textbf{0.929} & \textbf{0.854} & 0.593 & \textbf{0.849} & \textbf{0.872} & \textbf{0.998} \\
\bottomrule
\end{tabular}
\end{table}

The reading: gradient boosting on the engineered features is competitive on the chains that dominate the labelled pool --- close on Ethereum, at parity on Tron --- and falls 0.2--0.3 recall behind on Base and Arbitrum, \emph{with the same labels available to both models}. The graph's contribution concentrates exactly where a screening surface needs it: chains with thin label coverage. The producer script and the results artefact (sha-pinned) are committed alongside the other measurements.

\section{Deployment lessons}
\label{sec:lessons}

\textbf{Bit-exactness is cheap to demand and expensive to retrofit.} Requiring the deployed path to reproduce the offline artefact exactly (Sect.~\ref{sec:bitexact}) caught every class of silent divergence we encountered --- feature-order drift, quantisation differences, stale index shards --- at a cost of one comparison per release.

\textbf{Threshold provenance must travel with the threshold.} A threshold calibrated on one score carrier (a negative sample) and applied to another (the population) is off by about a bin. At a $10^{-3}$ budget over 323M addresses, a bin is thousands of alerts.

\textbf{Exact quantiles require the full pass.} Exact population quantiles over 835M nodes require scanning every score. The difference between an estimated $\tau$ and a measured $\tau$ is the difference between an alert volume an operator can state in advance and one they cannot.

\textbf{Evaluation snapshots are part of the threat model.} Sect.~\ref{sec:harness} Result 4 is not a story about simulators; it is a story about any evaluation whose sampling frame is correlated with the model under test.

\end{document}

%% file: event_table_appendix.tex
\begin{table}[!htbp]
\centering
\caption{The 42 Ethereum registry events of the replay (Sect.~\ref{sec:leadtime}). ``Date'' column: s = publication date scraped from the public source; o = OFAC action date from the registry date audit; p = proxy (event date + 3 days). ``Addr'' = in-graph / listed addresses; ``Flag'' = addresses flagged by the fincrime head at the $10^{-3}$ population budget; ``Lead'' = days from first on-chain appearance of the earliest flagged address to publication (--- = event not detected).}
\label{tab:events-ethereum}
\setlength{\tabcolsep}{3pt}
\tiny
\begin{tabular}{p{4.5cm}p{2.5cm}ccccr}
\toprule
Event & Category & Publication & Date & Addr & Flag & Lead (d) \\
\midrule
Suex OTC & sanctions\_designation & 2021-09-24 & p & 5/5 & 1 & -222.3 \\
Wormhole & bridge\_exploit & 2022-02-05 & p & 1/1 & 0 & --- \\
Ronin Bridge Exploit & bridge\_exploit & 2022-03-26 & p & 1/1 & 0 & --- \\
Ronin & bridge\_exploit & 2022-03-26 & p & 3/3 & 1 & 424.5 \\
OFAC SDN: GARANTEX EUROPE OU & ofac\_designation & 2022-04-05 & o & 9/9 & 0 & --- \\
OFAC SDN: LAZARUS GROUP & ofac\_designation & 2022-04-14 & o & 8/8 & 3 & -13.5 \\
Harmony Horizon Bridge & bridge\_exploit & 2022-06-27 & p & 5/5 & 1 & 528.8 \\
Nomad & bridge\_exploit & 2022-08-04 & p & 5/5 & 0 & --- \\
Tornado Cash & sanctions\_designation & 2022-08-11 & p & 5/5 & 1 & 98.7 \\
Euler Finance Exploit & defi\_exploit & 2023-03-14 & s & 1/1 & 0 & --- \\
Euler Finance & defi\_exploit & 2023-03-14 & s & 2/2 & 1 & 0.6 \\
OFAC SDN: SIM, Hyon Sop & ofac\_designation & 2023-04-24 & o & 9/9 & 2 & -788.2 \\
Multichain & bridge\_exploit & 2023-07-07 & s & 5/5 & 3 & 0.3 \\
OFAC SDN: SEMENOV, Roman & ofac\_designation & 2023-08-23 & o & 7/8 & 0 & --- \\
Mixin Network & exchange\_breach & 2023-09-25 & s & 4/4 & 2 & 2.0 \\
OFAC SDN: JIMENEZ CASTRO, Mario Alberto & ofac\_designation & 2023-09-26 & o & 1/1 & 0 & --- \\
Poloniex & exchange\_breach & 2023-11-10 & s & 5/5 & 3 & 1801.3 \\
KyberSwap & defi\_exploit & 2023-11-23 & s & 3/4 & 0 & --- \\
Orbit Bridge & bridge\_exploit & 2024-01-03 & s & 1/2 & 1 & 4.3 \\
Munchables & insider\_attack & 2024-03-27 & s & 5/5 & 0 & --- \\
WazirX & exchange\_breach & 2024-07-18 & s & 5/5 & 1 & 588.5 \\
Penpie & defi\_exploit & 2024-09-04 & s & 4/5 & 2 & 0.7 \\
BingX & exchange\_breach & 2024-09-20 & s & 5/5 & 2 & 0.1 \\
Radiant Capital Oct & defi\_exploit & 2024-10-17 & s & 3/5 & 0 & --- \\
OFAC SDN: MAGOMEDOV, Khadzhi Murat Dalgatovich & ofac\_designation & 2024-12-04 & o & 1/1 & 0 & --- \\
Orange Finance & defi\_exploit & 2025-01-09 & s & 1/5 & 0 & --- \\
Moby & key\_compromise & 2025-01-13 & s & 1/2 & 0 & --- \\
Phemex & exchange\_breach & 2025-01-24 & s & 5/5 & 3 & 1106.7 \\
ByBit Feb 2025 & exchange\_breach & 2025-02-22 & s & 2/2 & 0 & --- \\
North Korea Crypto Theft 2025 & sanctions\_designation & 2025-02-24 & p & 5/5 & 1 & 1026.7 \\
Infini & key\_compromise & 2025-02-24 & s & 4/4 & 0 & --- \\
OFAC SDN: FUNNULL TECHNOLOGY INC & ofac\_designation & 2025-05-29 & o & 2/2 & 0 & --- \\
OFAC SDN: OLD VECTOR LLC & ofac\_designation & 2025-08-14 & o & 2/2 & 0 & --- \\
OFAC SDN: DERAKHSHAN, Alireza & ofac\_designation & 2025-09-16 & o & 2/2 & 1 & 1370.5 \\
OFAC SDN: SOKOLOVSKI, Rolan & ofac\_designation & 2025-11-19 & o & 7/7 & 0 & --- \\
Genesis Crypto & sanctions\_designation & 2026-02-13 & s & 5/5 & 1 & 1380.7 \\
Hamas Wallets Round 1 & sanctions\_designation & 2026-02-13 & s & 5/5 & 1 & 1380.7 \\
Hamas Wallets Round 2 & sanctions\_designation & 2026-02-13 & s & 5/5 & 1 & 1380.7 \\
Fentanyl Precursor Wallets & sanctions\_designation & 2026-02-13 & s & 5/5 & 1 & 1380.7 \\
North Korea IT Worker Ring & sanctions\_designation & 2026-02-13 & s & 5/5 & 1 & 1380.7 \\
Cryptex CEX & sanctions\_designation & 2026-02-13 & s & 5/5 & 1 & 1380.7 \\
OFAC SDN: YUN, Song Guk & ofac\_designation & 2026-03-12 & o & 2/2 & 0 & --- \\
\bottomrule
\end{tabular}
\end{table}

\begin{table}[!htbp]
\centering
\caption{The 26 Tron registry events of the replay (Sect.~\ref{sec:leadtime}). ``Date'' column: s = publication date scraped from the public source; o = OFAC action date from the registry date audit; p = proxy (event date). ``Addr'' = in-graph / listed addresses; ``Flag'' = addresses flagged by the fincrime head at the $10^{-3}$ population budget; ``Lead'' = days from first on-chain appearance of the earliest flagged address to publication (--- = event not detected).}
\label{tab:events-tron}
\setlength{\tabcolsep}{3pt}
\tiny
\begin{tabular}{p{4.5cm}p{2.5cm}ccccr}
\toprule
Event & Category & Publication & Date & Addr & Flag & Lead (d) \\
\midrule
OFAC SDN: HANAFIN, John Desmond & ofac\_designation & 2023-05-19 & o & 1/1 & 0 & --- \\
OFAC SDN: SHAFIU, Ali & ofac\_designation & 2023-07-31 & o & 1/1 & 0 & --- \\
OFAC SDN: GAMBASHIDZE, Ilya Andreevich & ofac\_designation & 2024-03-20 & o & 2/2 & 0 & --- \\
OFAC SDN: AL-LAW, Tawfiq Muhammad Sa'id & ofac\_designation & 2024-03-26 & o & 1/1 & 1 & 412.5 \\
OFAC SDN: GAZA NOW & ofac\_designation & 2024-03-27 & o & 6/7 & 0 & --- \\
OFAC SDN: OBSHCHESTVO S OGRANICHENNOI OTVETSTVENNOSTYU KONSTRUKTORSKOE & ofac\_designation & 2024-08-23 & o & 1/1 & 0 & --- \\
OFAC SDN: CHIRKINYAN, Elena & ofac\_designation & 2024-12-04 & o & 1/1 & 0 & --- \\
OFAC SDN: AL-JAMAL, Sa'id Ahmad Muhammad & ofac\_designation & 2024-12-19 & o & 5/5 & 1 & 352.3 \\
Grinex Designation & sanctions\_designation & 2025-03-10 & p & 4/5 & 1 & 307.8 \\
Venezuela State Crypto Rails & sanctions\_designation & 2025-04-02 & p & 4/5 & 1 & 330.8 \\
OFAC SDN: ANSARALLAH & ofac\_designation & 2025-04-02 & o & 8/8 & 2 & 501.5 \\
OFAC SDN: AEZA GROUP LLC & ofac\_designation & 2025-07-01 & o & 1/1 & 0 & --- \\
OFAC SDN: GRINEX & ofac\_designation & 2025-08-14 & o & 7/7 & 0 & --- \\
OFAC SDN: ALIVAND, Arash Estaki & ofac\_designation & 2025-09-16 & o & 5/5 & 1 & 801.6 \\
OFAC SDN: SAYYED, Sadiq Abbas Habib & ofac\_designation & 2025-09-24 & o & 1/1 & 0 & --- \\
OFAC SDN: CHEIL CREDIT BANK & ofac\_designation & 2025-11-04 & o & 6/9 & 2 & 546.8 \\
OFAC SDN: WEDDING, Ryan James & ofac\_designation & 2025-11-19 & o & 3/3 & 1 & 568.2 \\
OFAC SDN: ZEDCEX EXCHANGE LTD & ofac\_designation & 2026-01-30 & o & 7/7 & 3 & 473.4 \\
Garantex & sanctions\_designation & 2026-02-13 & s & 4/5 & 1 & 647.8 \\
Russian Facilitation Ring & sanctions\_designation & 2026-02-13 & s & 4/5 & 1 & 647.8 \\
Cambodia Pig Butchering Ring & sanctions\_designation & 2026-02-13 & s & 4/5 & 1 & 647.8 \\
Iran IRGC Facilitators & sanctions\_designation & 2026-02-13 & s & 4/5 & 1 & 647.8 \\
Houthi Funding Network & sanctions\_designation & 2026-02-13 & s & 4/5 & 1 & 647.8 \\
Hamas Financial Facilitators 2024 & sanctions\_designation & 2026-02-13 & s & 4/5 & 1 & 647.8 \\
Iran UAV Procurement Crypto & sanctions\_designation & 2026-02-13 & s & 4/5 & 1 & 647.8 \\
OFAC SDN: AMNOKGANG TECHNOLOGY DEVELOPMENT COMPANY & ofac\_designation & 2026-03-12 & o & 6/6 & 2 & 913.7 \\
\bottomrule
\end{tabular}
\end{table}

%% file: paper-01-v2-arxiv.bbl
\begin{thebibliography}{38}

\bibitem{alarab2020}
Alarab, I., Prakoonwit, S., Nacer, M.I.: Competence of graph convolutional networks for anti-money laundering in Bitcoin blockchain. In: 5th International Conference on Machine Learning Technologies (ICMLT), pp. 23--27. ACM (2020)

\bibitem{altman2023aml}
Altman, E.R., Blanu\v{s}a, J., von Niederh\"ausern, L., Egressy, B., Anghel, A., Atasu, K.: Realistic synthetic financial transactions for anti-money laundering models. In: Advances in Neural Information Processing Systems 36 (NeurIPS 2023), Datasets and Benchmarks Track (2023)

\bibitem{androulaki2013privacy}
Androulaki, E., Karame, G.O., Roeschlin, M., Scherer, T., Capkun, S.: Evaluating user privacy in Bitcoin. In: Financial Cryptography and Data Security (FC 2013), pp. 34--51. Springer (2013)

\bibitem{apruzzese2023real}
Apruzzese, G., Anderson, H.S., Dambra, S., Freeman, D., Pierazzi, F., Roundy, K.A.: ``Real attackers don't compute gradients'': bridging the gap between adversarial ML research and practice. In: IEEE Conference on Secure and Trustworthy Machine Learning (SaTML), pp. 339--364 (2023)

\bibitem{arp2022dosdonts}
Arp, D., Quiring, E., Pendlebury, F., Warnecke, A., Pierazzi, F., Wressnegger, C., Cavallaro, L., Rieck, K.: Dos and don'ts of machine learning in computer security. In: 31st USENIX Security Symposium, pp. 3971--3988 (2022)

\bibitem{bellei2024elliptic2}
Bellei, C., Xu, M., Phillips, R., Robinson, T., Weber, M., Kaler, T., Leiserson, C.E., Arvind, Chen, J.: The shape of money laundering: subgraph representation learning on the blockchain with the Elliptic2 dataset. arXiv:2404.19109 (2024)

\bibitem{berner2019dota}
Berner, C., Brockman, G., Chan, B., et al.: Dota 2 with large scale deep reinforcement learning. arXiv:1912.06680 (2019)

\bibitem{chiang2019clustergcn}
Chiang, W.L., Liu, X., Si, S., Li, Y., Bengio, S., Hsieh, C.J.: Cluster-GCN: an efficient algorithm for training deep and large graph convolutional networks. In: 25th ACM SIGKDD International Conference on Knowledge Discovery and Data Mining (KDD), pp. 257--266 (2019)

\bibitem{daian2020flashboys}
Daian, P., Goldfeder, S., Kell, T., Li, Y., Zhao, X., Bentov, I., Breidenbach, L., Juels, A.: Flash Boys 2.0: frontrunning in decentralized exchanges, miner extractable value, and consensus instability. In: IEEE Symposium on Security and Privacy (S\&P), pp. 910--927 (2020)

\bibitem{egressy2024ppgn}
Egressy, B., von Niederh\"ausern, L., Blanu\v{s}a, J., Altman, E.R., Wattenhofer, R., Atasu, K.: Provably powerful graph neural networks for directed multigraphs. In: 38th AAAI Conference on Artificial Intelligence, pp. 11838--11846 (2024)

\bibitem{elmougy2023forensics}
Elmougy, Y., Liu, L.: Demystifying fraudulent transactions and illicit nodes in the Bitcoin network for financial forensics. In: 29th ACM SIGKDD Conference on Knowledge Discovery and Data Mining (KDD), pp. 3979--3990 (2023)

\bibitem{ferretti2025aml}
Ferretti, S., D'Angelo, G., Ghini, V.: Enhancing anti-money laundering frameworks: an application of graph neural networks in cryptocurrency transaction classification. IEEE Access \textbf{13}, 50201--50215 (2025)

\bibitem{gao2026ahgt}
Gao, J., Richard, B.S., Xia, H., Victor, K., Fabien, E.B., Xia, Q.: AHGT-DFD: adaptive hierarchical graph transformer for dynamic fraud detection in blockchain networks. IEEE Transactions on Dependable and Secure Computing \textbf{23}(2), 2229--2241 (2026)

\bibitem{guo2017calibration}
Guo, C., Pleiss, G., Sun, Y., Weinberger, K.Q.: On calibration of modern neural networks. In: 34th International Conference on Machine Learning (ICML), pp. 1321--1330 (2017)

\bibitem{hamilton2017}
Hamilton, W.L., Ying, R., Leskovec, J.: Inductive representation learning on large graphs. In: Advances in Neural Information Processing Systems 30 (NIPS 2017), pp. 1024--1034 (2017)

\bibitem{hausknecht2015}
Hausknecht, M., Stone, P.: Deep recurrent Q-learning for partially observable MDPs. In: AAAI Fall Symposium on Sequential Decision Making for Intelligent Agents, pp. 29--37. AAAI Press (2015)

\bibitem{hu2020hgt}
Hu, Z., Dong, Y., Wang, K., Sun, Y.: Heterogeneous graph transformer. In: The Web Conference (WWW), pp. 2704--2710 (2020)

\bibitem{huang2018ransomware}
Huang, D.Y., Aliapoulios, M.M., Li, V.G., Invernizzi, L., Bursztein, E., McRoberts, K., Levin, J., Levchenko, K., Snoeren, A.C., McCoy, D.: Tracking ransomware end-to-end. In: IEEE Symposium on Security and Privacy (S\&P), pp. 618--631 (2018)

\bibitem{li2022ttagn}
Li, S., Gou, G., Liu, C., Hou, C., Li, Z., Xiong, G.: TTAGN: temporal transaction aggregation graph network for Ethereum phishing scams detection. In: The Web Conference (WWW), pp. 661--669 (2022)

\bibitem{li2026dynamicphish}
Li, C., Zhang, Y., Liu, R., Zhao, Z., Duan, H., Zeng, Q.: Ethereum phishing fraud detection via dynamic graph neural network with anonymous walks. Pattern Recognition \textbf{180}, 114343 (2026)

\bibitem{lo2023inspection}
Lo, W.W., Kulatilleke, G.K., Sarhan, M., Layeghy, S., Portmann, M.: Inspection-L: self-supervised GNN node embeddings for money laundering detection in Bitcoin. Applied Intelligence \textbf{53}(16), 19406--19417 (2023)

\bibitem{lopezrojas2018paysim}
Lopez-Rojas, E.A., Axelsson, S., Baca, D.: Analysis of fraud controls using the PaySim financial simulator. International Journal of Simulation and Process Modelling \textbf{13}(4), 377--386 (2018)

\bibitem{meiklejohn2013}
Meiklejohn, S., Pomarole, M., Jordan, G., Levchenko, K., McCoy, D., Voelker, G.M., Savage, S.: A fistful of bitcoins: characterizing payments among men with no names. In: Proceedings of the 2013 Internet Measurement Conference (IMC), pp. 127--140. ACM (2013)

\bibitem{pareja2020evolvegcn}
Pareja, A., Domeniconi, G., Chen, J., Ma, T., Suzumura, T., Kanezashi, H., Kaler, T., Schardl, T.B., Leiserson, C.E.: EvolveGCN: evolving graph convolutional networks for dynamic graphs. In: 34th AAAI Conference on Artificial Intelligence, pp. 5363--5370 (2020)

\bibitem{pendlebury2019tesseract}
Pendlebury, F., Pierazzi, F., Jordaney, R., Kinder, J., Cavallaro, L.: TESSERACT: eliminating experimental bias in malware classification across space and time. In: 28th USENIX Security Symposium, pp. 729--746 (2019)

\bibitem{qin2022bev}
Qin, K., Zhou, L., Gervais, A.: Quantifying blockchain extractable value: how dark is the forest? In: IEEE Symposium on Security and Privacy (S\&P), pp. 198--214 (2022)

\bibitem{ron2013quantitative}
Ron, D., Shamir, A.: Quantitative analysis of the full Bitcoin transaction graph. In: Financial Cryptography and Data Security (FC 2013), pp. 6--24. Springer (2013)

\bibitem{schulman2017}
Schulman, J., Wolski, F., Dhariwal, P., Radford, A., Klimov, O.: Proximal policy optimization algorithms. arXiv:1707.06347 (2017)

\bibitem{velickovic2018gat}
Veli\v{c}kovi\'{c}, P., Cucurull, G., Casanova, A., Romero, A., Li\`{o}, P., Bengio, Y.: Graph attention networks. In: 6th International Conference on Learning Representations (ICLR) (2018)

\bibitem{vinyals2019}
Vinyals, O., Babuschkin, I., Czarnecki, W.M., et al.: Grandmaster level in StarCraft II using multi-agent reinforcement learning. Nature \textbf{575}(7782), 350--354 (2019)

\bibitem{wang2026pdhg}
Wang, L., Mi, Y., Zhang, Y., Zhang, J.: PDHG: an Ethereum phishing detection approach via heterogeneous graph transformer. Expert Systems with Applications \textbf{298}, 129919 (2026)

\bibitem{wang2023mixers}
Wang, Z., Chaliasos, S., Qin, K., Zhou, L., Gao, L., Berrang, P., Livshits, B., Gervais, A.: On how zero-knowledge proof blockchain mixers improve, and worsen user privacy. In: The Web Conference (WWW), pp. 2022--2032 (2023)

\bibitem{weber2018scalable}
Weber, M., Chen, J., Suzumura, T., Pareja, A., Ma, T., Kanezashi, H., Kaler, T., Leiserson, C.E., Schardl, T.B.: Scalable graph learning for anti-money laundering: a first look. arXiv:1812.00076 (2018)

\bibitem{weber2019}
Weber, M., Domeniconi, G., Chen, J., Weidele, D.K.I., Bellei, C., Robinson, T., Leiserson, C.E.: Anti-money laundering in Bitcoin: experimenting with graph convolutional networks for financial forensics. In: KDD Workshop on Anomaly Detection in Finance (2019), arXiv:1908.02591

\bibitem{wu2022phishers}
Wu, J., Yuan, Q., Lin, D., You, W., Chen, W., Chen, C., Zheng, Z.: Who are the phishers? Phishing scam detection on Ethereum via network embedding. IEEE Transactions on Systems, Man, and Cybernetics: Systems \textbf{52}(2), 1156--1166 (2022)

\bibitem{ying2018pinsage}
Ying, R., He, R., Chen, K., Eksombatchai, P., Hamilton, W.L., Leskovec, J.: Graph convolutional neural networks for web-scale recommender systems. In: 24th ACM SIGKDD International Conference on Knowledge Discovery and Data Mining (KDD), pp. 974--983 (2018)

\bibitem{yousaf2019tracing}
Yousaf, H., Kappos, G., Meiklejohn, S.: Tracing transactions across cryptocurrency ledgers. In: 28th USENIX Security Symposium, pp. 837--850 (2019)

\bibitem{zeng2020graphsaint}
Zeng, H., Zhou, H., Srivastava, A., Kannan, R., Prasanna, V.K.: GraphSAINT: graph sampling based inductive learning method. In: 8th International Conference on Learning Representations (ICLR) (2020)

\end{thebibliography}
